\documentclass[preprint2]{emulateapj-rtx4}

\slugcomment{To appear in the ApJ}

\usepackage{txfonts} 
\usepackage{natbib} 
\bibpunct{(}{)}{;}{a}{}{,}
\newcommand{\kms}{~{\rm km~s}^{-1}}
\newcommand{\Msun}{M_\odot}
\newcommand{\nth} {n_{\rm thr}}
\newcommand{\pcc} {{\rm ~cm}^{-3}}
\newcommand{\psc} {{\rm ~cm}^{-2}}
\newcommand{\racc} {r_{\rm acc}}

\begin{document}

\title{From dusty filaments to massive stars: The case of NGC 7538 S}
 
%\title{Is NGC 7538 S a massive disk or a fragmenting filament?}

\author{Raul Naranjo-Romero\altaffilmark{1}, Luis A. Zapata\altaffilmark{1}, Enrique V\'azquez-Semadeni\altaffilmark{1}, Satoko Takahashi\altaffilmark{2},  
Aina Palau\altaffilmark{3}, and Peter Schilke\altaffilmark{5} }

\altaffiltext{1}{Centro de Radioastronom{\'\i}a y Astrof{\'\i}sica, Universidad 
Nacional Aut\'onoma de M\'exico, Morelia 58090, Mexico}
\altaffiltext{2}{Academia Sinica Institute of Astronomy and Astrophysics, P.O. Box 23-141, Taipei 10617, Taiwan}
\altaffiltext{3}{Institut de Ci{\`e}ncies de l'Espai (CSIC-IEEC), Campus UAB-Facultat de Ci{\`e}ncies, Torre C5-parell 2, 08193, Bellaterra, Spain}
\altaffiltext{5}{I. Physikalisches Institut,  Universitat zu Koln, Zulpicher Strasse 77, 50937, Koln, Germany}

\begin{abstract}
We report on high-sensitivity and high-angular resolution archival
Submillimeter Array (SMA) observations of the large ( $\sim$15000 AU) 
putative circumstellar disk associated with the O-type protostar NGC 7538 S. 
Observations of the continuum resolve this putative circumstellar
disk into five compact sources, with sizes $\sim$ 3000 AU and masses
$\sim10 M_\odot$. This confirm the results of recent millimeter observations
made with CARMA/BIMA towards this object. 
However, we find that from most of these compact 
sources eject collimated bipolar outflows, revealed by our silicon monoxide (SiO {\it
J}=5-4) observations and confirm that these sources have a (proto)stellar nature.  
All outflows are perpendicular to the large and rotating dusty structure. 
We propose therefore that, rather than being a single massive
circumstellar disk, NGC 7538 S could be instead a large and massive
contracting or rotating filament that
is fragmenting at scales of 0.1 to 0.01 pc to form several B-type stars, via
the standard process involving outflows and disks. As in recent high
spatial resolution studies of dusty filaments, our observations also
suggest that thermal pressure does not seem to be sufficient to support
the filament, so that either additional support
needs to be invoked, or else the filament must be in the process of
collapsing. An SPH numerical simulation of the formation of a molecular
cloud by converging warm neutral medium flows produces contracting
filaments whose dimensions and spacings between the stars forming within
them, as well as their column densities, strongly resemble those observed in the 
filament reported here.
\end{abstract}


\section{Introduction}

The observational investigation of the formation of young and very
massive stars (with L$_*$ $\geq$ 10$^5$ $M_\odot$, {\it i.e.} O-type)
has been inherently a difficult task. These young stars are rare and
found far away from us, at typical distances of more than 2 kpc. Their
evolutionary time scales are also much shorter than those found in low
and intermediate mass stars \citep{mck2002}. At present, there is only a
handful of very massive star forming regions that have been extensively
studied at infrared, (sub)millimeter, and/or radio wavelengths using
sensitive, high angular resolution observations. Some of them include
G240.31+0.07 \citep{qiu2009}, IRAS 16547-4247 \citep{garay,ramon}, W51
North \citep{zapata2009,zapata2010}, NGC 7538 IRS1 \citep{qiu2011}, and
G24 A1 \citep{beltran2011}. These young O-type stars are found in
association with large ($\sim$ 5000 AU) accreting rotating structures
that seem to energize powerful bipolar outflows suggesting that these
stars may form in a similar way to the low mass stars, {\it i.e.} via
outflows and disks. However, it is very likely that those large accreting structures are forming 
more than one single high-mass star at once. These structures then are not 
disks in the sense of accretion disks that funnel gas to a single star.  

Here, we present high angular resolution archival observations using the
Submillimeter Array at millimeter wavelengths towards the extremely
large accreting structure surrounding the O-type protostar NGC 7538 S.
This is a cold core with a size of 0.5 pc, and a dust mass
of $\sim$ 2000 $M_\odot$ associated with H$_2$O, OH and CH$_3$OH (but
not SiO) maser emission, and weak free-free emission
\citep{san2004,pes2006,zapata2009b}.  This region is at a distance of
2.65 $\pm$ 0.12 kpc \citep{mosca2009}.  In the middle of this cold core
resides a large ($\sim$ 15~000 AU) and massive ($\sim$ 400
$M_\odot$) structure interpreted as a nearly
edge-on rotating disk that surrounds a protostar with 
a mass between 20 -- 40 M$_\odot$ \citep{san2003,san2010}. 
However, the values reported by \citet{san2010}, which should be the most recent estimates, 
are significantly lower, 85-115 M$_\odot$ for the putative disk, 
and 14-24 M$_\odot$ for the enclosed mass of the central protostar and the inner, 
rotationally supported part of the disk.

A single, young, and powerful outflow perpendicular to the rotating disk has also
been mapped in molecular lines that includes SiO ({\it J}=2-1) and
HCO$^+$({\it J}=1-0) \citep{san2010}.  The bipolar outflow is compact
(0.2 pc), has a low velocity ($\Delta v$ $\sim$ 40 km s$^{-1}$), and has its
blueshifted side towards the northwest, and the redshifted one  to the
southeast.

Very recent BIMA and CARMA observations have resolved the 
large and massive disk into four compact millimeter sources (S$_{A1,A2,B,C}$) 
\citep{mel2012, cor2008}.  Whether the massive and unstable disk fragmented in four pieces or 
whether a dusty core fragmented into four protostars is unclear. 

\begin{table*}
\begin{center}
\scriptsize
\caption{Observational parameters of the SMA compact sources}
\begin{tabular}{lcccccc}
\tableline\tableline &\multicolumn{2}{c}{Position$^a$} &  Flux Density & Flux Peak &  Deconvolved Angular Size$^b$ & dust-gas mass$^c$\\ 
\cline{2-3}     Name$^d$       & $\alpha$(J2000) & $\delta$(J2000) &  (mJy) &  (mJy) & &  ($M_\odot$) \\ 
\tableline 
SMA1 (CARMA S$_C$) &  23 13 44.483 & $+$61 26 47.70 &  130$\pm$30  & 41$\pm$4 & $1\rlap.{''}5 \pm 0\rlap.{''}6 \times 1\rlap.{''}2 \pm 0\rlap.{''}8;~  78^\circ \pm 40^\circ$&  10.0\\ 
SMA2 (CARMA S$_B$) &  23 13 44.770 & $+$61 26 48.85 &  110$\pm$26  & 44$\pm$4 & $1\rlap.{''}3 \pm 0\rlap.{''}4 \times 0\rlap.{''}4 \pm 0\rlap.{''}5;~  19^\circ \pm 53^\circ$&  8.0\\
SMA3 (CARMA S$_{A1}$) &  23 13 44.979 & $+$61 26 49.41 &  143$\pm$34  & 56$\pm$3 & $1\rlap.{''}5 \pm 0\rlap.{''}5 \times 1\rlap.{''}3 \pm 0\rlap.{''}6;~  45^\circ \pm 35^\circ$&  13.0\\
SMA4 (CARMA S$_{A2}$) &  23 13 45.099 & $+$61 26 50.04 &    -   &  -  & - &  -\\
SMA5  -- &  23 13 45.267 & $+$61 26 50.47 &    80$\pm$30  & 12$\pm$3 &  $1\rlap.{''}6 \pm 0\rlap.{''}9 \times 1\rlap.{''}0 \pm 0\rlap.{''}8;~  133^\circ \pm 44^\circ$& 7.0\\
 \tableline
\end{tabular}
\tablenotetext{a}{Units of right ascension are hours, minutes, and
 seconds and units of declination are degrees, arcminutes, and
 arcseconds.}  \tablenotetext{b}{Major axis $\times$ minor axis;
 position angle of major axis. The values were obtained 
 using the task JMFIT of AIPS.}
 \tablenotetext{c}{The values of the masses obtained here have
uncertainties of order a factor of 2 due to the error in
the estimation of temperatures of the millimeter sources and the error in the dust mass opacity coefficient at this wavelength.
See main text for dust temperature and dust opacity assumptions.}
 \tablenotetext{d}{The CARMA names were obtained from \citet{mel2012,cor2008}.}
\end{center}
\end{table*}

\begin{figure}[ht]
\begin{center}
\includegraphics[scale=0.29]{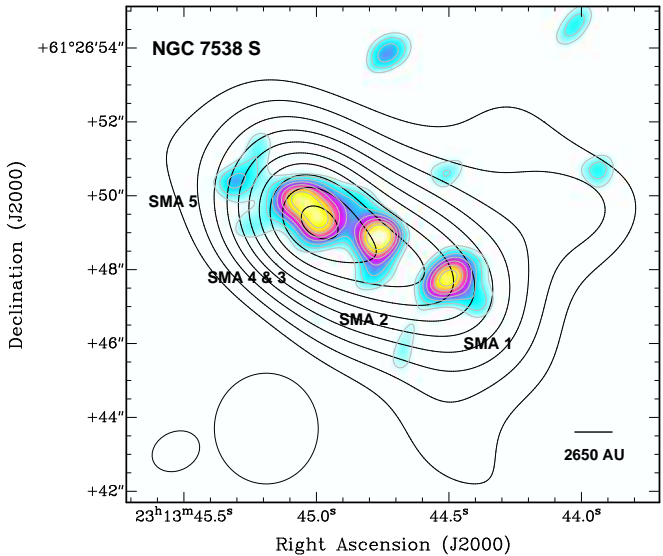}
\caption{\scriptsize Color and grey contour map of the 1.3 mm continuum emission from NGC 7538 S region 
obtained with the extended configuration and overlaid
 in black contours with the millimeter continuum image obtained with the compact configuration.  
 The grey contours range from 20\% to 90\% with steps of 10\% of the peak of the emission. 
 The black contours range from 55\% to 90\% with steps of 7\% of the peak of the emission.
 The synthesized beams of the 1.3 mm continuum observations are described
 in the text and are shown in the bottom left corner of the image.}
\label{fig1}
\end{center}
\end{figure}

\section{Observations}

The observations were obtained from the SMA\footnote{The
Submillimeter Array (SMA) is a joint project between the Smithsonian
Astrophysical Observatory and the Academia Sinica Institute of Astronomy
and Astrophysics, and is funded by the Smithsonian Institution and the
Academia Sinica.} archive, and were made on october 2005 and october
2006, when the array was respectively in its compact and extended configurations.  The
independent baselines in the compact and extended configurations ranged
in projected length from 5 to 50 k$\lambda$ and 15 to 120 k$\lambda$,
respectively.  The phase reference center for the observations was at
R.A. = 23h13m44.46s, decl.= $+$61$^\circ$26$'$49.8$''$ (J2000.0).  The
frequency was centered at 217.238539 GHz in the Lower Sideband (LSB),
while the Upper Sideband (USB) was centered at 227.238539 GHz. The
primary beam of the SMA in the neighborhood of 220 GHz has a FWHM
$\sim 50''$. The emission from the whole NGC 7538 S region falls well
inside of the FWHM. The largest brightness distribution to which these
observations are sensitive enough (more than the 50\% level) is $\sim$
27$''$ in FWHM when the distribution is Gaussian \citep{wil1994}.

The SMA digital correlator was configured in 24 spectral
windows (``chunks'') of 104 MHz each, with 128 channels distributed
over each spectral window, providing a resolution of 0.812 MHz ($\sim$ 1.106 km
s$^{-1}$) per channel. 
 
The zenith opacity ($\tau_{230 GHz}$), measured with the NRAO tipping
radiometer located at the Caltech Submillimeter Observatory, was between
0.1 and 0.2, indicating reasonable weather conditions during the
observations. Observations of Uranus provided the absolute scale for the
flux density calibration.  The gain calibrators were the quasars BL LAC
and 0102+584.  The uncertainty in the flux scale is estimated to be
between 15 and 20$\%$, based on the SMA monitoring of quasars.  Further
technical descriptions of the SMA and its calibration schemes can be
found in \citet{Hoetal2004}.

The data were calibrated using the IDL superset MIR, originally
developed for the Owens Valley Radio Observatory 
\citep[OVRO,][]{Scovilleetal1993} and adapted for the SMA.\footnote{The MIR-IDL
cookbook by C. Qi can be found at
http://cfa-www.harvard.edu/$\sim$cqi/mircook.html} The calibrated data
were imaged and analyzed in the standard manner using the MIRIAD and
KARMA \citep{goo96} softwares.  We set the ROBUST parameter of the task
INVERT to 0 to obtain an optimal compromise between resolution and
sensitivity. The resulting rms noise for the continuum high resolution
image was around 10 mJy beam$^{-1}$ at an angular resolution of
$1\rlap.{''}33$ $\times$ $1\rlap.{''}04$ with a P.A. = $-63.8^\circ$,
while for the low resolution image it was around 5 mJy beam$^{-1}$ at an
angular resolution of $3\rlap.{''}23$ $\times$ $3\rlap.{''}15$ with a
P.A. = $6.0^\circ$.

For the line emission, the resulting synthesized beam was $1\rlap.{''}65$ $\times$ $1\rlap.{''}37$ with a P.A. =
$-69.8^\circ$ (SiO). The rms noise for the line image was about 150
mJy.  We concatenated the uv-data from both configurations to make 
the line emission map. 
%The difference in the angular
%resolution between the two maps is due to the fact that in the SiO map
%we set the ROBUST parameter of the task INVERT (MIRIAD) to $-$2 to
%obtain a slightly better resolution sacrificing some sensitivity.  
It is important to mention that in these observations many more lines were
detected. However, in this work we concentrate only on the molecular
distribution of the SiO.  This molecule is well known to be
good tracer of high density gas in outflows with critical densities of
more than 10$^{3}$ cm$^{-3}$.

 \begin{figure*}[ht]
\begin{center}\bigskip
\includegraphics[scale=0.49]{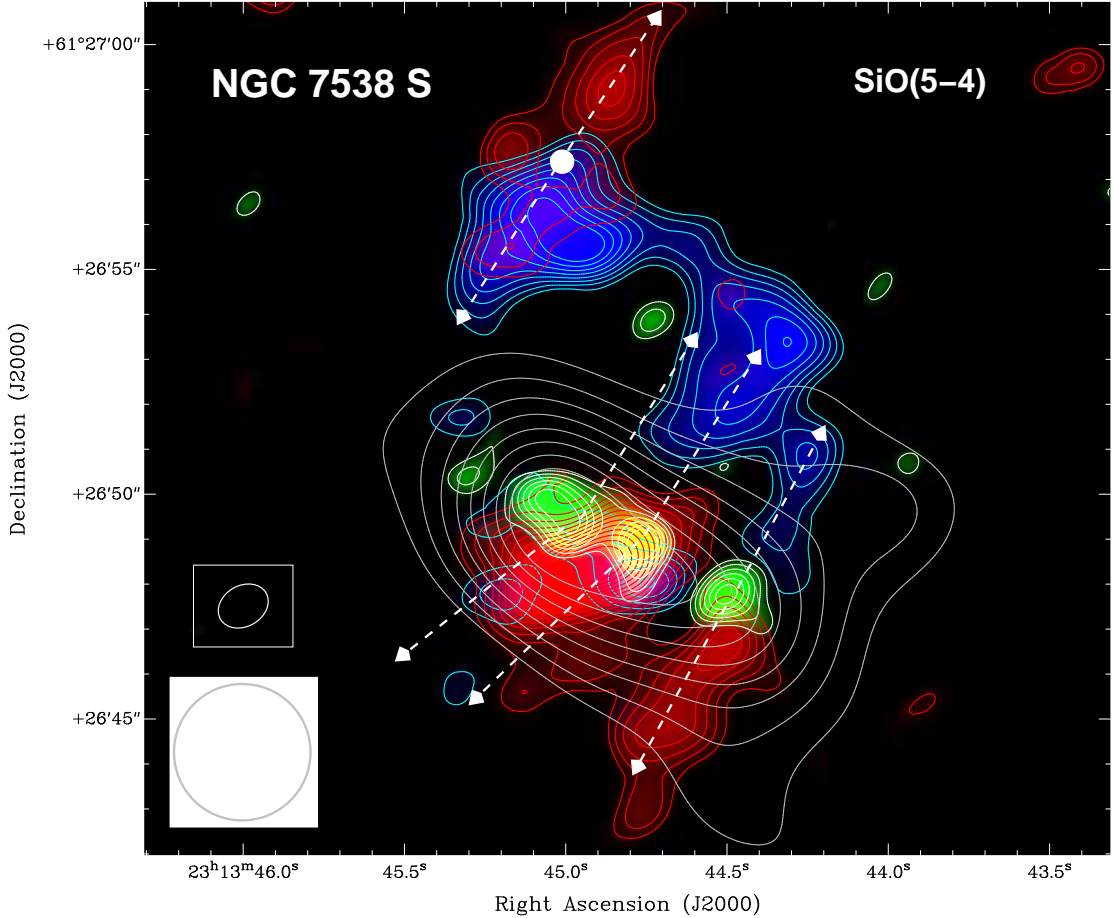}
\caption{\scriptsize Integrated intensity color and contour map of the
SiO ({\it J}=5-4) line emission 
observed toward NGC 7538 S overlaid in contours and green color with the SMA continuum emission 
from Figure 1.  The blue and red contours are 
 from 30\% to 90\% with steps of 7\% of the peak of the line emission; 
 the peak the SiO({\it J}=5-4) emission is 100 Jy Beam$^{-1}$ km s$^{-1}$. 
 The white contours are from 30\% to 90\% with steps of 7\% of
 the peak of the emission; the peak of the 1.3 mm map
 is 730 mJy Beam$^{-1}$. The green contours are from 20\% to 10\% with steps of 7\% 
 of the peak of the emission. The synthesized beams of the SiO and 1.3 mm continuum images are described
 in the text and are shown in the left lower corner of the figure. 
 The white dot marks the position of the infrared source found in the
archive of the Spitzer Space Telescope.  The white arrows mark the orientation
of the collimated outflows revealed in the SiO (J=5-4) observations.
} 
\label{fig2}
\end{center}
\end{figure*}

\section{RESULTS}

\subsection{Continuum emission}

In Figure 1, we show the resulting millimeter ($\lambda$ = 1300 $\mu$m)
continuum image obtained with the SMA in the compact and
extended configurations from the high-mass star forming region NGC 7538
S. From the compact configuration, we detect a single
northeast-southwest extended dusty source reported for the first time by
\citet{san2003,san2010} and interpreted as an extremely large edge-on
circumstellar disk.  However, \citet{mel2012}  made BIMA and CARMA observations 
and resolved the large and massive disk into four compact millimeter sources (S$_{A1,A2,B,C}$) 
\citep{mel2012}. The nature of these source is still in debate. 

This source has a deconvolved size of $9\rlap.{''}4
\pm 0\rlap.{''}6 \times 5\rlap.{''}1 \pm 0\rlap.{''}3$ with a position
angle $58^\circ \pm 8^\circ$. Its total flux density at this wavelength
is 4.1 $\pm$ 0.5 Jy.  On the other hand, in the extended configuration,
we resolved the elongated dusty source into five compact sources
(SMA1-5), with the compact objects being located along the major axis
of the source detected with the compact configuration.  These compact
sources have sizes of about 1.5$''$ (or 3000 AU at the distance of NGC
3875 S, see Table 1). The deconvolved parameters of the sources SMA3/4 were 
determined by fitting a two-dimensional Gaussian function to the observed flux distribution
using the task JMFIT of AIPS\footnote{Astronomical Image Processing
System (AIPS) of the NRAO: http://www.aips.nrao.edu/index.shtml}. 
SMA3 was well resolved using this method while for the fainter source SMA4 failed. 
Thus, we only give the position on the sky for SMA4. A better estimation of the 
deconvolved sizes of SMA3/4 (CARMA S$_{A1}$ and S$_{A2}$) are given in \citet{mel2012}.
  
Four of these sources (SMA1, 2, 3 and 4) coincide well with the positions of
the four compact sources found by \citet{cor2008, san2010, mel2012} using
observations of CARMA/BIMA, which they propose to certainly have a
protostellar nature.  The source SMA 5 does not have any 
counterpart at mm wavelengths as revealed by the Figure 8 of  \citet{mel2012}.
We are thus caution about if this source is real or is an artifact by poor the uv-sampling. 
However, as our synthesized beam is larger
to the ones obtained by the CARMA and BIMA observations in \citet{mel2012},
perhaps, we are only recovering some more extended emission and revealing   
the non-compact sources.  The density fluxes for the sources SMA1, 2, 3 and 4 that we 
obtained here were larger (by a factor of more than 3) than those obtained by \citet{mel2012}, 
again probably because we are recovering more extended emission. 
The deconvolved sizes are in good agreement with those presented in \citet{mel2012}.

Following \citet{zapata2011}, we adopted a value of $\kappa_{\rm 1.3 mm}$ =
0.1 cm$^2$ g$^{-1}$ \citep{oss1994} and assumed optically thin, isothermal dust
emission and a gas-to-dust ratio of 100, with a dust temperature of 30 K
\citep[]{san2010} for these millimeter compact objects, deriving dust
masses of about 10 solar masses (see Table 1). From our compact
configuration observations, we estimate that the elongated
northeast-southwest dusty structure reported by \citet{san2003} and
\citet{san2010} has a mass of 350 $M_\odot$, which is in very good
agreement with the mass obtained by those authors.
The assumption of a dust temperature of 30 K appears justified, see, e.g., 
\citet[]{san2004,san2010}. 
For example, Zheng et al. (2001) find ammonia to be optically thick in the direction 
of NGC 7538 S with a temperature of 25 K, however, the low resolution
of the observations did not allow to measure the temperatures of every single mm
component of NGC 7538 S. 

We noted that \citet{mel2012} obtained a mass of 60 M$_\odot$ for the source
CARMA S$_{A1}$ or SMA3. This value is very different from that we obtained (about 10 M$_\odot$).
This discrepancy could arise from the low value of $\kappa_{\rm 3 mm}$ = 0.002 cm$^2$ g$^{-1}$ 
they used. There is not a general consensus of using a singular value, however,  
this dust mass opacity at this wavelength seems to be too low..

\subsection{Line emission}

%\subsubsection{Molecular outflows}

In our 2 GHz lower side band of the observations, we detected the line
SiO ({\it J}=5-4) at a rest frequency of 217.104 GHz. In Figure 2, we
present a map of the intensity, integrated over velocity (moment 0), of
the SiO line emission observed towards NGC7538S. The map is additionally
overlaid with the two maps of the continuum emission obtained from the
compact and extended configurations (Figure 1).  The velocity
integration is over the velocity ranges: $-$70 to $-$57 km s$^{-1}$
(blue) and $-$53 to $-$40 km s$^{-1}$ (red). This map reveals at least
two collimated and bipolar outflows emanating from NGC 7538 S, all with
similar orientations. The bipolar outflows emanate with their
blueshifted sides toward the northwest while the redshifted sides to the
southeast. These outflows have position angles of approximately
$-$45$^\circ$ and with collimation angles of
less than 10$^\circ$. The extension of the outflows is about 0.1 pc.  

There is one more bipolar
outflow, revealed here for the first time, which is located about 7$''$
north of NGC 7538 S (Figure 2). However, we do not find any millimeter
continuum source in the outflow center.  The outflow is also very
collimated and with a similar orientation to those found in NGC 7538
S. An infrared study with the archive images of the Spitzer Infrared
Telescope revealed an infrared (3.6 $\micron$) faint source in the
middle of the outflow, whose position is R.A. =
23h13m45.121s, decl.=$+$61$^\circ$26$'$57.01$''$ (J2000.0). We suggest
that this infrared source could be its exciting source.  For the rest of
the outflows in the field, the compact millimeter sources reported here
(SMA1, 2 and 3) appear to be exciting them.  We also find an IRAC (4.5
and 5.8 $\micron$) source associated with the millimeter compact source
SMA3 \& 4.  The infrared emission detected towards this object might
suggest a more evolved evolutionary state compared with the rest of the
compact millimeter sources.  However, the nature of this infrared source
is uncertain and an in-deep infrared study of the SMA3 \& 4 is beyond of
the scope of the present Letter.

The collection of collimated outflows reported here appear to be part of
the massive, bipolar, unresolved outflow reported by
\citet{san2003} and \citet{san2010}. This has similar position angle
($+$148$^\circ$), range of radial velocities ($-$80, $-$40 km s$^{-1}$),
and projected size (25$''$ or 0.2 pc).

The compact source SMA3/4 is additionally associated with a radio thermal
jet with an orientation similar to the molecular outflows
\citep{san2003, san2010, mel2012} also suggesting that this might be the exciting
source of another outflow.

Following \citet{zapata2009}, and assuming local thermodynamic equilibrium (LTE), 
that the molecular emission is optically thin, an excitation temperature of 50 K,
and an abundance ratio of SiO/H$_2$ equal to 1 $\times$ 10$^{-7}$, we
can estimate the mass of the outflows for the SiO molecule in the
transition {\it J} = 5 - 4. We obtain a mass of about 10 $M_\odot$ for
each outflow. It is important to mention that the fractional abundance
between the silicon monoxide and the molecular hydrogen varies from one
star-forming region (SFR) to another, and thus the mass of the outflow
can be over/underestimated. The value for the fractional abundance
assumed here seems to be consistent for a few SFRs \citep[see][]{ziu1987,
zha1995}.

For a mechanical force of F$_M$ = 10 $M_\odot$ 10 km s$^{-1}$/10,000 yr
= 0.01 $M_\odot$ km s$^{-1}$ yr$^{-1}$, and from the correlation
presented in \citet{wu2004} for the outflow mechanical force versus the
bolometric luminosity of the exciting source, we very roughly estimate a
luminosity for the central powering sources of the order of 10$^{3}$
L$_\odot$, which corresponds to a massive B-type protostar. This
spectral type for the central star is in good agreement with that
obtained from the dust emission, as we estimated on the last section.

 \begin{figure*}[ht]
\begin{center}\bigskip
\includegraphics[scale=0.37, angle=0]{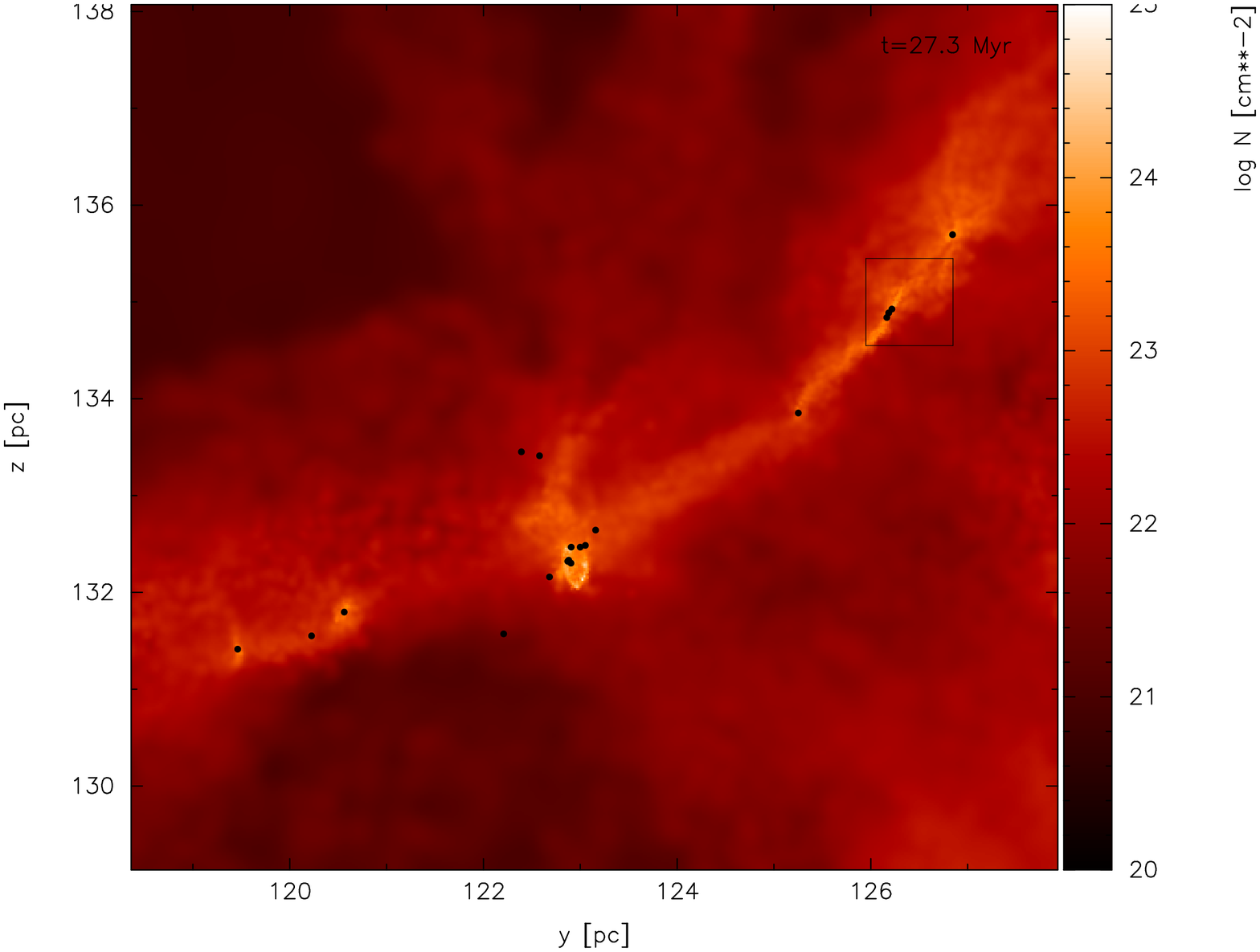}
\includegraphics[scale=0.37, angle=0]{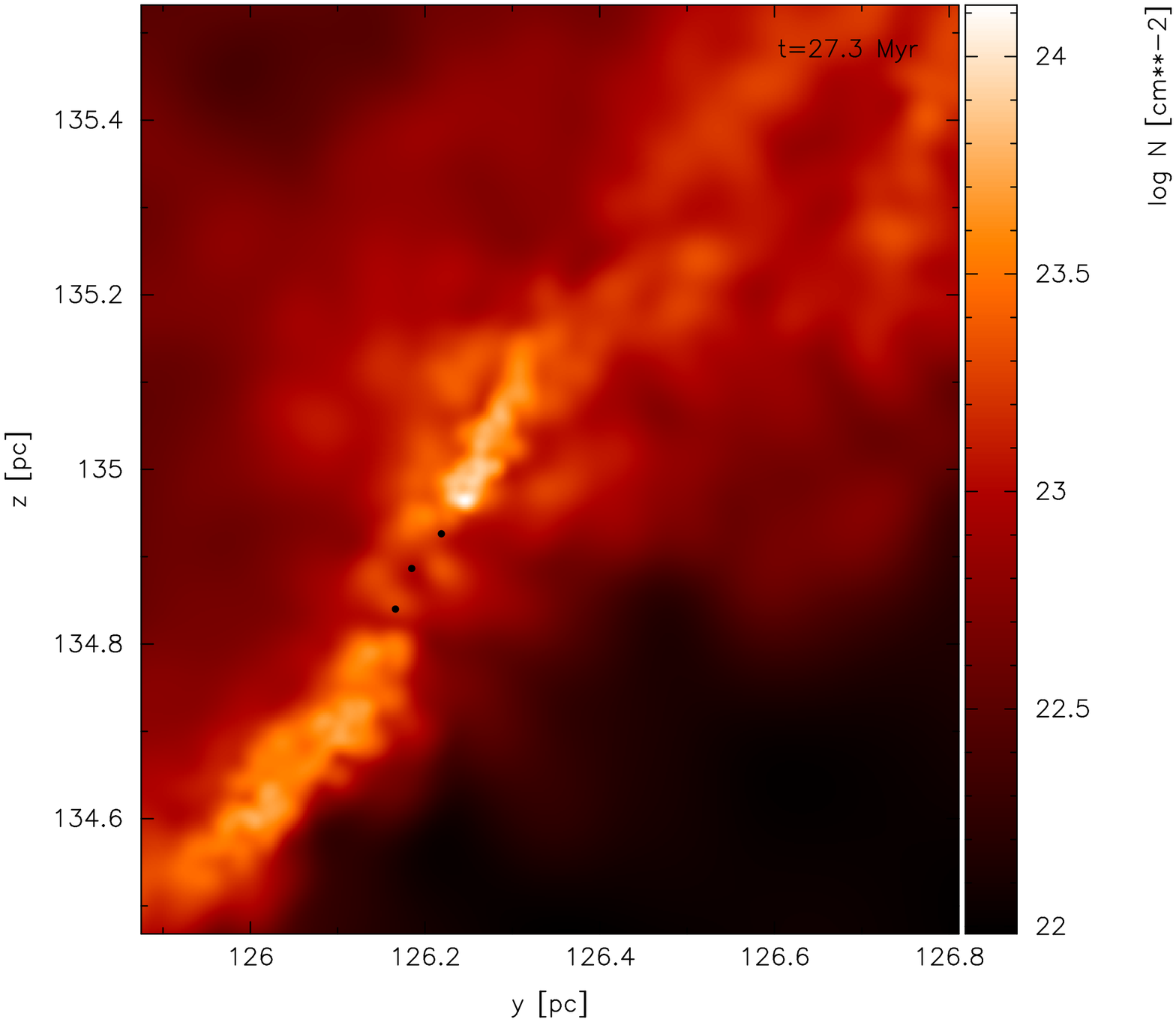}
\caption{\scriptsize Column density in a numerical simulation of
molecular cloud formation by converging warm neutral medium flows, in a
region where a filament develops by gravitational contraction. The right
panel shows an enlargement of the region delimited by the square in the
left panel. Note the different column density ranges for the two
panels. See the text for a more complete discussion of the figures. }
\label{fig3}
\end{center}
\end{figure*}

\section{Numerical Simulation}

Figure 4 shows a filament forming in a new numerical simulation of
molecular cloud formation by converging warm neutral medium (WNM) flows,
similar to the one presented by \citet{VS+07}, but with 8 times higher
mass resolution. The simulation will be described in detail in a
separate contribution, but here we give the basic details. It was
performed with the SPH code Gadget 2 \citep{Springel05}, with a
prescription for ``sink'' particle creation first applied by
\citet{Jappsen+05}.  According to this prescription, when the local
density $n$ at any location within the simulation exceeds a certain
threshold density $\nth$, the SPH gas particles within that region are
replaced by a single sink particle, whose mass and momentum equal
the sums of the masses and momenta of the removed gas particles. After
being created, the sink particles behave as point masses, and can
continue to accrete gas from a region of radius $\racc$ around them, if
such gas is gravitationally bound to the sinks.

The present simulation contains $296^3 \approx 2.6 \times 10^7$ SPH
particles, and starts at a volume density $n = 1 \pcc$ and $T = 5000$ K,
in a numerical box of 256 pc per side. The total mass in the box is thus
$5.23 \times 10^5 \Msun$. In this box, we set up as initial velocity
conditions two cylindrical gas streams of radius 32 pc and length 112
pc, about to collide head-on at the plane $x = 128$ pc (i.e., at the
middle of the box), with velocity $9.2 \kms$ \citep[for more details on
this setup, see][]{VS+07}. For the sink particle parameters, we choose
$\nth = 3.2 \times 10^7 \pcc$ and $\racc = 0.04$ pc.

The left panel of Fig. 4 shows the column density at time 27.3 Myr in a
$9 \times 9$ pc region on the $(y,z)$ plane, near the center of the
simulation, with a density integration over the range $118 \le x \le
130$ pc. In this region, a long filament has formed by gravitational
contraction, with several star-forming sites along it.

The square in the left panel of Fig.\ 4 delimits the region shown in the
right panel. This panel shows the smaller-scale filament arising within
the larger one, immediately after forming three sink (stellar)
particles, with masses $\sim 3 \Msun$ each. 

The similarity between the filament+sinks system in the simulation and 
NGC7538 S is striking. The length spanned by the three sinks, as seen in 
the right panel of Fig.\  3, is $\sim 0.1$ pc, while the corresponding
length in NGC 7538 S is $\sim 6"$, corresponding to $0.08$ pc. The 
sinks in the simulation, as well as the individual sources in NGC 7538 S are
almost evenly spaced. The width 
of the filament containing the sinks in the simulation is also of approximately
the same size, as is the width of the dust-emitting region in the source (Fig.\ 1). 
We can also estimate the column density of NGC 7538 S, and compare
it to that of the filament in the simulation. The estimated mass of the region is 
$350 \Msun$. Assuming that it has the same extension along the line of sight
as it is wide on the plane of the sky ($\sim 0.08$ pc), and a cylindrical shape, 
with length $\sim 9" \sim 0.12$ pc, the resulting column density is $N \sim 
10^{24} \psc$, which compares very well with the range of column densities 
for the filament in the simulation (right panel of Fig.\ 3). 

Finally, it is worth remarking that NGC 7538 S is located within a larger filament
spanning at least $3' $  \citep[corresponding to $\sim 2.4$ pc;][]{san2010}, while the filament from the 
simulation shown in the
right panel of Fig.\ 3 can be seen in the left panel
of the same figure to be part of a larger filament, spanning also $\sim 2$ pc (the
filament containing the box in this figure).

Note, however, that, in the simulation, this filament is in a state of gravitational
contraction, both accreting mass in the direction perpendicular to its
axis as well as contracting along this axis. Nevertheless, the simulated
filament is seen to strongly resemble the observed one in both
size scale and spacing of the newly-formed stars.

\section{Discussion and Conclusions}

The dimensions ($\sim$ 3000 AU), masses (of about 10 $M_\odot$), and the
fact that from most of the sources molecular collimated outflows or
thermal jets emerge, suggest that those compact sources are probably large
envelopes or perhaps dusty large disks associated with massive stars in
the process of formation. 
We thus conclude that NGC 7538 S is not likely to be a single, young O-type
star surrounded by an extremely large disk, but rather maybe a 
large contracting or rotating filament that
is fragmenting at scales of 0.1 pc to 0.01 pc 
to form multiple high-mass stars that involve outflows, envelopes and disks.
Similar molecular and dusty fragmenting filaments have already been
reported in the literature in OMC3 and G28.34+0.06 \citep{sato2006,
wang2011}. For the case of G28.34+0.06, observations also from the SMA
revealed a string of five dust cores of along the 1 pc IR-dark filament
called G28.34+0.06. The cores are well aligned at a position angle of
$+$48$^\circ$, and are regularly spaced, with an average projected separation of 0.16 pc. For
OMC3, there is a large filament that also fragments
at scales of 0.01 pc to form the the protostellar object MM7 \citep{sato2006}. 
This large filament also shows large velocity gradients along it.  
This case is very similar to the case of NGC 7538 S reported here.

Following \citet{Pil2011} and using the density derived from the
emission detected in the compact configuration (1.3 $\times$ 10$^7$
cm$^{-3}$), a velocity dispersion with a thermal line width (assuming 30
K) of 0.77 km s$^{-1}$ (dispersion of 0.33 s$^{-1}$), we estimate a
Jeans mass of M$_{\rm J}$ = 0.76 $M_\odot$ and a Jeans length of
$\lambda_{\rm J}$ = 1930 AU. Thus, for our case, the Jeans mass is
smaller than the measured masses (by one order of magnitude, as in G28),
and the Jeans length is slightly smaller than the measured separation
between the fragments. So we also conclude, in agreement with
\citet{wang2008, wang2011}, that thermal pressure does not seem to be
enough to support the source, which appears to be a filament, so that
either additional support needs to be invoked, or else that the filament
is in the process of gravitational contraction, as is the case of the
filament in the numerical simulation presented here.

Finally, we note that our results are consistent with recent BIMA, CARMA
and SPITZER observations presented in \citet{mel2012} and confirm the
presence of multiple compact source located along of NGC7538 S.

\acknowledgments 

R.N.R., E.V.S., and L.A.Z. acknowledge the financial
support from DGAPA, UNAM, and CONACyT, M\'exico. 
A.P. is supported by
the Spanish MICINN grant AYA2008-06189-C03 (co-funded
with FEDER funds) and by a JAE-Doc CSIC fellowship
co-funded with the European Social Fund.
The numerical simulation was performed on the compute cluster 
acquired with funding from grant CONACYT 102488, to E.V.-S.

\end{document}